\input{aipcheck}

\documentclass[
    ,final            
  ]
  {aipproc}

\layoutstyle{6x9}

\begin{document}

\title{Solar neutrino detection}


\author{Lino Miramonti}{
  address={Physics department of Milano University and INFN}
}

\begin{abstract}
More than 40 years ago, neutrinos where conceived as a way to test the validity of the solar models which tell us that stars are powered by nuclear fusion reactions. The first measurement of the neutrino flux, in 1968 in the Homestake mine in South Dakota, detected only one third of the expected value, originating what has been known as the Solar Neutrino Problem.
Different experiments were built in order to understand the origin of this discrepancy. Now we know that neutrinos undergo oscillation phenomenon changing their nature traveling from the core of the Sun to our detectors.
In the work the 40 year long saga of the neutrino detection is presented; from the first proposals to test the solar models to last real time measurements of the low energy part of the neutrino spectrum.
\end{abstract}

\maketitle


\section{How the Sun shines and neutrinos production}
The core of the Sun reaches temperatures of about 1.5 million K. At these temperatures, fusion can occur which transforms 4 protons into 1 helium nucleus. This latter has a mass that is slightly (0.7\%) smaller than the combined mass of the 4 protons. The \textit{missing mass} is converted into energy to power the Sun. Being the mass of the sum of the 4 protons equals to $6.6943 \cdot 10^{-27}$ kg and the mass of 1 helium nucleus equals to $6.6466 \cdot 10^{-27}$ kg, using $E=mc^2$ we found that each fusion releases $4.3 \cdot 10^{-12}$ J (or $26.7$ MeV). 
The current luminosity of the Sun is $4 \cdot 10^{26}$ W. 
\cite{Bahcall}

About 600 million tons of hydrogen per second is being converted to 596 million tons of helium. The remaining 4 million tons is released as energy.

The reaction starts from 4 protons and it end with 1 He nucleus which is composed of 2 protons and 2 neutrons.
This means that we have to transform 2 protons into 2 neutrons: $p \rightarrow n + e^+ + \nu_e$. When a proton becomes a neutron, an electron neutrino $\nu_e$ is released. Since neutrinos only interact with matter via the weak force, neutrinos generated by solar fusion pass immediately out of the core and into space. 
The study of solar neutrinos was conceived as a way to test the nuclear fusion reactions at the core of the Sun. \cite{BahcallandDavis}

\section{Solar neutrinos as a way to test the nuclear fusion reactions at the core of the Sun}
\begin{figure}
  \includegraphics[height=.28\textheight]{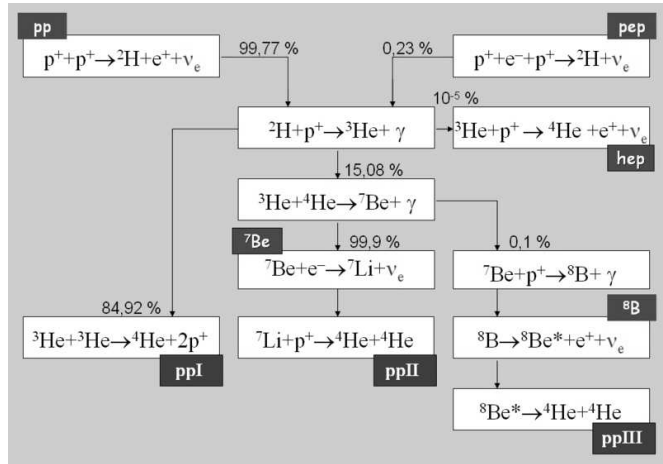}
  \caption{The pp chain reaction}
  \label{fig:The pp chain reaction}
\end{figure}

\begin{figure}
  \includegraphics[height=.35\textheight]{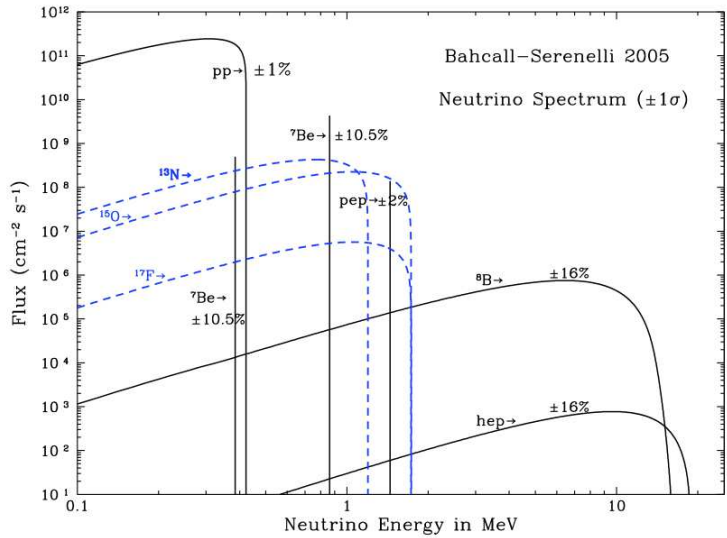}
  \caption{Neutrino energy spectrum as predicted by the Solar Standard Model (SSM)}
  \label{fig:Neutrino energy spectrum}
\end{figure}

In our star $\approx 98\%$ of the energy is created in the pp chain reaction. There are different steps in which energy (and neutrinos) are produced (see figure \ref{fig:The pp chain reaction}).
Beside pp chain reaction there is also the CNO cycle that become the dominant source of energy in stars heavier than the Sun. In our star the CNO cycle represents only 1-2$\%$.
Figure \ref{fig:Neutrino energy spectrum} shows the neutrino energy spectrum as predicted by the Solar Standard Model(SSM).

\subsection{Homestake: The first solar neutrino detector}
The first experiment built to detect solar neutrinos was performed by Raymond Davis and John N. Bahcall in the late 60's in the Homestake mine in South Dakota. It consisted in a large tank containing 615 tons of liquid perchloroethylene. Neutrinos are detected via the reaction: $\nu_e + ^{37}Cl \rightarrow ^{37}Ar + e^-$. The energy threshold is $E_{th} = 814$ keV and the most of the detected neutrinos are $^8B$ neutrinos.
\cite{Homestake}

The $^{37}Ar$ is radioactive and decays, with a half live of 35 days, via the reaction
$^{37}Ar + e^- \rightarrow  ^{37}Cl^* + \nu_e$.
The expected rate was of only 1 atom of $^{37}Ar$ every six days in 615 tons $C_2Cl_4$. 
About 5 atoms of $^{37}Ar$ were extracted  per month bubbling helium through the tank.
A new unit, the Solar Neutrino Unit (SNU), was introduced. A SNU correspond to 1 capture/sec/$10^{36}$ atoms. 

The expected number from SSM was of $7.6 + 1.3 - 1.1$ SNU while the detected one in Homestake was $2.56 \pm± 0.23$ SNU.
The number of neutrino detected was about 1/3 lower than the number of neutrino expected.

\subsection{The rise of Solar Neutrino Problem (SNP)}
This deficit induced physicists to speak about the Solar Neutrino Problem (SNP for short).
There are different possibilities to try to explain this puzzle. The first one is to admit that the Standard Solar Model is not correct, but solar models have been tested independently by helioseismology (that is the science that study the interior of the Sun by looking at its vibration modes), and the Standard Solar Model has so far passed all the tests. Besides non-standard solar models seem very unlikely.
The second possibility is that the Homestake experiment is wrong; for this reason since the beginning of '80 new experiments, employing different techniques, were built.
The last possibility is that something happens to the neutrinos.

\subsection{The real time detection of solar neutrinos: Kamiokande and SuperKamiokande}
Kamiokande and SuperKamiokande are large water Cherenkov Detectors; the first one is composed of 3000 tons of pure water viewed by 1000 PMTs, the second one is big version made of 50000 tons of pure water and 11200 PMTs.
The detected reaction is the elastic scattering on electron: $e^- + \nu \rightarrow e^- + \nu$. The energy thresholds are $E_{th} = 7.5$ MeV for Kamiokande and $E_{th} = 5.5$ MeV for SuperKamiokande and the detected neutrinos are mostly coming from $^8B$ and hep neutrinos.
The inferred flux was $\approx$ 2 times lower than the one predicted; confirming that the experiments detect less neutrinos that expected.
\cite{KamiokandeandSK}

\subsection{Looking for pp neutrinos: Gallex/GNO and SAGE}
Until the beginning of '90 there was no observation of the initial reaction in the nuclear fusion chain (i.e. pp neutrinos \footnote{these neutrinos are less model-dependent.}). This changed with the installation of the gallium experiments. Gallium as target allows neutrino interaction via: $\nu_e + ^{71}Ga \rightarrow ^{71}Ge + e^-$ featuring an energy threshold of only $E_{th} = 233$ keV.
Two radiochemical experiments were built in order to detect solar pp neutrinos. The GALLEX experiment (then GNO), located in the underground Gran Sasso laboratory (LNGS) in Italy, is composed of a tank contained 30 tons of natural gallium in a 100 tons of aqueous gallium chloride solution and the SAGE experiment located in the Baksan underground laboratory in Russia which is made of 50 tons of metallic gallium.
Gallex/GNO and SAGE measured a neutrino signal that was smaller than predicted by the solar model (of $\approx 60\%$).
Both experiment underwent calibration tests with an artificial neutrino source ($^{51}Cr$) confirming the proper performance of the detector.
\cite{Galiumexp}

All the experiments detect less neutrino than expected from the SSM.

\section{Neutrino oscillations}
The Standard Model of particle physics assumes that neutrinos are massless; let us assume that neutrinos have different masses $\Delta m^2$ and let us assume that the mass eigenstates ($\nu_1, \nu_2, \nu_3$), in which neutrinos are created and detected, are different form flavor eigenstates ($\nu_e,\nu _\mu, \nu_\tau$). We can write (in the simple case of 2 $\nu$):

\[
\left( {\begin{array}{*{20}c}
   {{\rm \nu }_{\rm \mu } }  \\
   {{\rm \nu }_{\rm e} }  \\
\end{array}} \right) = \left( {\begin{array}{*{20}c}
   {{\rm cos\theta }} & {{\rm sin\theta }}  \\
   { - {\rm sin\theta }} & {{\rm cos\theta }}  \\
\end{array}} \right)\left( {\begin{array}{*{20}c}
   {{\rm \nu }_{\rm 1} }  \\
   {{\rm \nu }_{\rm 2} }  \\
\end{array}} \right)
\]

Being:
\[
{\rm \nu }_{\rm i} \left( {\rm t} \right) = {\rm \nu }_{\rm i} \left( {\rm 0} \right){\rm e}^{ - {\rm iE}_{\rm i} {\rm t}} 
\]

In general this leads to the disappearance of the original neutrino flavour

\[
P(\nu _e  \to \nu _\mu  ) = \sin ^2 2\theta \sin ^2 \frac{{\Delta m^2 }}{{4E}}L
\]

with the corresponding appearance of the wrong neutrino flavour. The oscillation length is $L_{osc} = 4 \pi E/ \Delta m^2$. \cite{oscill}


The survival probability P depends upon two experimental parameters; the distance L from neutrino source to detector (in km) and the energy E (in GeV) and two fundamental parameters that are $\Delta m^2 = m_1^2 - m_2^2$ (in $eV^2$) and $sin^2 2\theta$ (where $\theta$ analogous to the Cabibbo angle in case of quarks).

In the real case of three neutrino we have\

\[
\left( {\begin{array}{*{20}c}
   {\nu _e }  \\
   {\nu _\mu  }  \\
   {\nu _\tau  }  \\
\end{array}} \right) = U\left( {\begin{array}{*{20}c}
   {\nu _1 }  \\
   {\nu _2 }  \\
   {\nu _3 }  \\
\end{array}} \right)
\]

where

\[
\begin{array}{*{20}c}
   {U = \left( {\begin{array}{*{20}c}
   1 & 0 & 0  \\
   0 & {{\rm cos}\theta _{{\rm 23}} } & {{\rm sin}\theta _{{\rm 23}} }  \\
   0 & {{\rm  - sin}\theta _{{\rm 23}} {\rm  }} & {{\rm cos}\theta _{{\rm 23}} }  \\
\end{array}} \right) \times \left( {\begin{array}{*{20}c}
   {{\rm cos}\theta _{{\rm 13}} } & 0 & {{\rm  sin}\theta _{{\rm 13}} e^{ - i\delta } }  \\
   0 & 1 & 0  \\
   {{\rm  -  sin}\theta _{{\rm 13}} e^{i\delta } } & 0 & {{\rm cos}\theta _{{\rm 13}} }  \\
\end{array}} \right) \times }  \\
   { \times \left( {\begin{array}{*{20}c}
   {{\rm cos}\theta _{{\rm 12}} } & {{\rm sin}\theta _{{\rm 12}} } & 0  \\
   {{\rm  - sin}\theta _{{\rm 12}} } & {{\rm cos}\theta _{{\rm 12}} } & 0  \\
   0 & 0 & 1  \\
\end{array}} \right) \times \left( {\begin{array}{*{20}c}
   {\exp (i\alpha_1)} & 0 & 0  \\
   0 & {\exp (i\alpha_2)} & 0  \\
   0 & 0 & 1  \\
\end{array}} \right)}  \\
\end{array}
\]

U is the Pontecorvo-Maki-Nakagawa-Sakata matrix and is the analog of the CKM matrix in the hadronic sector of the Standard Model ($\delta$ is the CP violation Dirac phase and $\alpha_1$ and $\alpha_2$ the CP-violating Majorana phases).

Neutrino oscillations can be enhanced by traveling through matter. This effect is knows as Mikheyev-Smirnov-Wolfenstein Effect (MSW) or matter effect. \cite{SMA} The neutrino \textit{index of refraction} depends on its scattering amplitude with matter; the Sun is made of up/down quarks and electrons so all neutrinos can interact through NC equally and only electron neutrinos can interact through CC scattering. For this reason the \textit{index of refraction} seen by $\nu _e$ is different than the one seen by $\nu _\mu$ and $\nu _\tau$.
The MSW effect gives for the probability of an $\nu _e$ produced at $t=0$ to be detected as a $\nu _\mu$: 

\[
\begin{array}{*{20}c}
   {P(\nu _e  \to \nu _\mu  ) = \sin ^2 2\theta _m \sin ^2 (\frac{{\pi xW}}{{\lambda _{osc} }})} & {where} & \begin{array}{l}
 \sin ^2 2\theta _m  = \frac{{\sin ^2 2\theta }}{{W^2 }} \\ 
 W^2  = \sin ^2 2\theta  + (D - \cos ^2 2\theta )^2  \\ 
 D = \sqrt 2 G_F N_e \frac{{2E_\nu  }}{{\Delta m^2 }} \\ 
 \end{array}  \\
\end{array}
\]
$N_e$ being the electron density.

\section{The Sudbury Neutrino Observatory}
The Sudbury Neutrino Observatory (SNO), located in a mine at a depth of 6010 m of water equivalent, is composed of 1000 tons of heavy water contained in a 12 m diameter acrylic vessel viewed by 9500 PMTs. \cite{SNO}
Thanks to the reaction on deuteron SNO is able to look for different neutrino reactions:
the elastic scattering (ES) reaction $\nu_x + e^- \rightarrow \nu_x + e^-$,
the neutral current (NC) reaction $\nu_x + d \rightarrow p + n + \nu_x$ and 
the charged current (CC) reaction $\nu_e + d \rightarrow p + p + e^-$.
\
Charged and neutral current fluxes are measured independently and the results are:
\
\[
\begin{array}{l}
 \phi _{CC}  = 1.68{\rm   }_{ - 0.06}^{ + 0.06} ({\rm stat}{\rm .})_{ - 0.09}^{ + 0.08} ({\rm syst}{\rm .}) \\
 \phi _{NC}  = 4.94{\rm   }_{ - 0.21}^{ + 0.21} ({\rm stat}{\rm .})_{ - 0.34}^{ + 0.38} ({\rm syst}{\rm .}) \\ 
 \phi _{ES}  = 2.35{\rm   }_{ - 0.22}^{ + 0.22} ({\rm stat}{\rm .})_{ - 0.15}^{ + 0.15} ({\rm syst}{\rm .}) \\ 
 \end{array}
\]
in units of $10^6\ cm^{-2} s^{-1}$.
The total flux of active neutrinos is measured independently (NC) and agrees well with the Standard Solar Model calculations: $4.7 \pm 0.5$ (BPS07).
The survival probability is 
\[
P_{ee}  = \frac{{\phi _{CC} }}{{\phi _{NC} }} = 0.34 \pm 0.023({\rm stat}{\rm .})_{ - 0.03112}^{ + 0.029} 
\]

\subsection{Solution of the Solar Neutrino Problem}

\begin{figure}
  \includegraphics[height=.3\textheight]{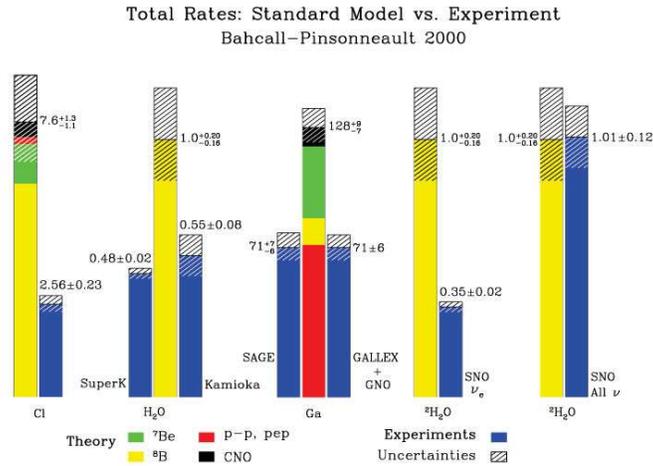}
  \caption{The total rates Solar Standard Model vs the experiment}
  \label{fig:Solar Standard Model vs the experiment}
\end{figure}

All experiments \textit{see} less neutrinos than expected by SSM but SNO in case of NC. Figure \ref{fig:Solar Standard Model vs the experiment} shows the total rates expected from the Solar Standard Model vs the experiments. The solution of the SNP is that electron neutrinos produced in the core of the Sun oscillate into non-electron neutrino with the following parameters: $\Delta m_{12}^2  = 7.6 \cdot 10^{ - 5} eV^2$ and $\sin ^2 2\vartheta _{12}  = 0.87$.

These parameters correspond to the Large Mixing Angle (LMA) region in the frame of MSW matter effect.
These values are obtained combining all the solar neutrino experiments and the one obtained by the KamLAND experiment. \footnote{KamLAND is a detector built to measure electron antineutrinos coming from 53 commercial power reactors (average distance of $\approx 180$ km ). The experiment is sensitive to the neutrino mixing associated with the(LMA) solution.}

\section{Real time measurement below 1 MeV: Borexino}
All the water solar experiments described so far are able to detect in real time only neutrinos having an energy greater than some MeV \footnote{Corresponding to only about 0.01\% of the total flux.}. The radiochemical detectors (i.e. Homestake, Gallex/GNO and SAGE) integrate over time and energy. The Borexino detector, located in the Gran Sasso laboratory in central Italy, is able to measure for the first time, neutrino coming from the Sun in real time with low energy ($\approx$ 200 keV) and high statistic looking for elastic scattering (ES) on electrons in a very high purity liquid scintillator (active volume of $\approx$ 100 tons). In order to reach such a low energy threshold extreme radiopurity of the scintillator has to be obtained (the amount of $^{238}U$ and $^{232}Th$ do not exceded  $10^{-16} g(U)/g$ and $g(Th)/g$ equivalent). \cite{Arpesella}

The main goal of Borexino is the measurement of  $^7Be$ neutrinos, thank to that it will be possible to test the Standard Solar Model and 
to test the fundamental prediction of MSW-LMA theory in the regime transition, expected between 1-2 MeV, between the vacuum driven oscillations and the matter effect oscillations. Moreover it will be possible to provide a strong constraint on the $^7Be$ rate, at or below 5\%, such as to provide an  essential input to check the balance between photon luminosity and neutrino luminosity of the Sun.

\subsection{The $^7Be$ neutrino flux measurement}
Borexino starts the data taking with the detector completely filled in May 2007.
The measured amount\footnote{Here the data refers to 192 days of data taking.} of $^{238}U$ is $1.6 \pm± 0.1 \cdot 10^{-17} g(U)/g$ corresponding to 1.9 cpd/100 tons, and the measured amount of $^{232}Th$ is $6.8 \pm± 1.5 \cdot 10^{-18} g(Th)/g$ corresponding to 0.25 cpd/100 tons. \cite{Borexino192day}
For each event the time and the total charge are measured, besides, the position of each event is reconstructed with algorithms based on time of flight fit to hit time distribution of detected photoelectrons; this allows to define a fiducial volume (of about 100 tons of effective mass) in order to maximize the signal to noise ratio (the spatial resolution of reconstructed events is about 16 cm at 500 keV). The light yield has been evaluated fitting the $^{14}C$ spectrum and the $^{11}C$ spectrum and is equal to $500 \pm 12 photoelectrons/MeV$. The energy resolution is $5\%$ at 1 MeV (scaling as $5\%/\sqrt{E\left[MeV\right]}$).

\begin{figure}
  \includegraphics[height=.2\textheight]{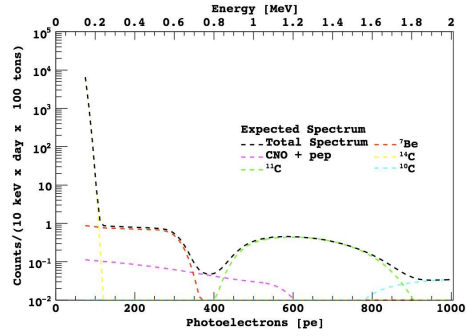}
  \caption{The expected spectrum in Borexino. In Abscisse the number of photoelectrons and in ordinate the counts per day in 100 tons of fiducial volume}
  \label{fig:expected_spectrum}
\end{figure}

\begin{figure}
  \includegraphics[height=.2\textheight]{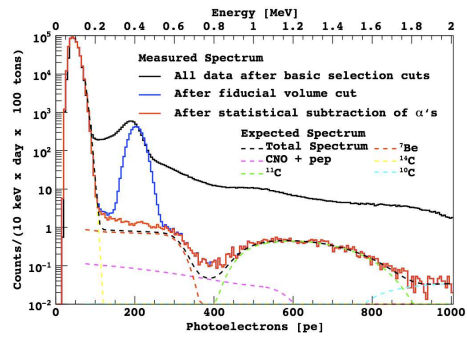}
  \caption{The obtained spectrum in Borexino (see text). In Abscisse the number of photoelectrons and in ordinate the counts per day in 100 tons of fiducial volume}
  \label{fig:final_result}
\end{figure}

Figure \ref{fig:expected_spectrum} shows the expected spectrum. The final results are presented in figure \ref{fig:final_result}; the black line are the raw data collected in 192 day of data taking before any rejection cuts; the blue line is obtained keeping only the events inside the fiducial volume and the red line is obtained after removing the alpha particles with the pulse shape discrimination. Fitting the obtained spectrum the measured $^7Be$ flux is $49 \pm 3_{sys} \pm 4_{stat}$ cpd/100 tons to be compared with the expected $^7Be$ interaction rate for MSW-LMA oscillations that is $48 \pm 4$ cpd/100 tons in the case of high metallicity and $44 \pm 4$ cpd/100 tons for the low metallicity \footnote{Abundance of all elements above helium}.

\section{Perspectives and conclusions}
More then forty years have elapsed from the first idea to test the solar models trying to detect neutrinos produced in the termonuclear reactions powering the Sun. The deficit observed by Raymond Davis was the hint that neutrinos change their nature traveling from the core of the Sun to our detectors. Thank to 40 years of experiments now we know that neutrino oscillates and we understand quite well how the Sun, and the other stars, shines. In the next few years the Borexino detector will reveal to us, which of the two models (high or low metallicty) is the correct one.

\begin{theacknowledgments}
I wish to thank the organizers of the III School on Cosmic Rays and Astrophysics held in Arequipa in August 2008; in particular I wish to thank Jose Bellido and Javier Solano which invited me to write this contribution. I also thank John Michael who helped me in writing this work.
\end{theacknowledgments}

\bibliographystyle{aipproc}   

\end{document}